\documentstyle[11pt,newpaspZI,twoside]{article}
\def\edcomment#1{\iffalse\marginpar{\raggedright\sl#1\/}\else\relax\fi}
\reversemarginpar

\begin{document}

\newcommand\x         {\hbox{$\times$}}
\newcommand\othername {\hbox{$\dots$}}
\def\eq#1{\begin{equation} #1 \end{equation}}
\def\eqarray#1{\begin{eqnarray} #1 \end{eqnarray}}
\def\eqarraylet#1{\begin{mathletters}\begin{eqnarray} #1 %
                  \end{eqnarray}\end{mathletters}}
\def\mic              {\hbox{$\mu{\rm m}$}}
\def\about            {\hbox{$\sim$}}
\def\Mo               {\hbox{$M_{\odot}$}}
\def\Lo               {\hbox{$L_{\odot}$}}
\def\comm#1           {{\tt (COMMENT: #1)}}
\def\u                {\hbox{$u^*$}}
\def\g                {\hbox{$g^*$}}
\def\r                {\hbox{$r^*$}}
\def\i                {\hbox{$i^*$}}
\def\z                {\hbox{$z^*$}}
\def\ug               {\hbox{$u^*-g^*$}}
\def\ur               {\hbox{$u^*-r^*$}}
\def\gr               {\hbox{$g^*-r^*$}}
\def\ri               {\hbox{$r^*-i^*$}}
\def\gi               {\hbox{$g^*-i^*$}}
\def\iz               {\hbox{$i^*-z^*$}}
\def\Fpeak            {\hbox{F$_{peak}$}}
\def\Fint             {\hbox{F$_{int}$}}
\def\t                {\hbox{$t$}}

\title{           The Optical, Infrared and Radio Properties of   
        Extragalactic Sources Observed by SDSS, 2MASS and FIRST Surveys }

\author{\v{Z}. Ivezi\'c$^1$, R.H. Becker$^2$, M. Blanton$^3$, X. Fan$^1$, 
K. Finlator$^1$, J.E. Gunn$^1$, P. Hall$^1$, R.S.J. Kim$^4$, G.R. Knapp$^1$,  
J. Loveday$^5$, R.H. Lupton$^1$, K. Menou$^1$, V. Narayanan$^1$, G.R. Richards$^6$, 
C.M. Rockosi$^7$, D. Schlegel$^1$, D.P. Schneider$^6$, I. Strateva$^1$, 
M.A. Strauss$^1$, D. Vanden Berk$^8$, W. Voges$^9$, B. Yanny$^8$, for the SDSS 
Collaboration}

\affil{
 $^1$Princeton University, 
 $^2$University of California,
 $^3$The New York University, 
 $^4$The John Hopkins University, 
 $^5$University of Sussex,
 $^6$Pennsylvania State University, 
 $^7$University of Washington,
 $^8$Fermi National Accelerator Laboratory,
 $^9$Max-Planck-Institute f\"ur Extraterrestrische Physik
}

\begin{abstract}
We positionally match sources observed by the Sloan Digital Sky Survey 
(SDSS), the Two Micron All Sky Survey (2MASS), and the Faint Images
of the Radio Sky at Twenty-cm (FIRST) survey. Practically all 2MASS sources 
are matched to an SDSS source within 2 arcsec; $\sim$11\% of them are optically 
resolved galaxies and the rest are dominated by stars. About 1/3 of FIRST 
sources are matched to an SDSS source within 2 arcsec; $\sim$80\% of these 
are galaxies and the rest are dominated by quasars. Based on these results, 
we project that by the completion of these surveys the matched samples will 
include about 10$^7$ stars and 10$^6$ galaxies observed by both SDSS and 
2MASS, and about 250,000 galaxies and 50,000 quasars observed by both
SDSS and FIRST. Here we present a preliminary analysis of the optical, infrared 
and radio properties for the extragalactic sources from the matched samples. 
In particular, we find that the fraction of quasars with stellar colors 
missed by the SDSS spectroscopic survey is probably not larger than \about10\%, 
and that the optical colors of radio-loud quasars are \about 0.05 mag. redder
(with $4\sigma$ significance) than the colors of radio-quiet quasars.   
\end{abstract}

\section{Introduction}

The increasing availability of large scale digital sky surveys spanning
many wavelengths offers an unprecedented view of the Universe.
The positional matching of such surveys is of obvious scientific interest. 
Not only can wide wavelength coverage provide a comprehensive description 
of the various classes of astrophysical object, but characterizing the most 
populous families can help isolating more peculiar, and usually more 
interesting, objects. In this contribution we discuss the matching of early 
SDSS data with 2MASS and FIRST surveys. The details of this work are presented 
elsewhere (Finlator {\em et al.} 2000, Menou {\em et al.} 2001, Knapp {\em 
et al.} 2001, Ivezi\'{c} {\em et al.} 2001, Ivezi\'{c} {\em et al.} 2002) and 
here we summarize the most important results. 

The SDSS (York  {\em et al.} 2000, Stoughton {\em et al.} 
2002, and references therein) is a digital photometric and spectroscopic survey 
which will cover one quarter of the Celestial Sphere in the North Galactic cap and 
produce a smaller area ($\sim$225 deg$^2$) but much deeper survey in the Southern
Galactic hemisphere. It utilizes five broad bands ($u'$, $g'$, $r'$, $i'$, $z'$) with 
central wavelengths 
ranging from 3550~\AA\ to 8930~\AA, and will detect about 10$^8$ stars and a similar 
number of galaxies brighter than $\sim$22$^m$. For about one million brightest galaxies 
and 100,000 quasar candidates SDSS will also obtain high-quality spectra. 
The 2MASS (Skrutskie {\em et al.} 1997) surveyed the entire 
sky in near-infrared light ($J$, $H$, and $K_s$ bands) and catalogued $\sim$300 
million stars, as well as several million galaxies. The FIRST survey
(Becker {\em et al.} 1995) provides 
the most  comprehensive view of the Universe at 20 cm and will detect about a 
million radio galaxies and quasars.

\begin{figure}[t]
\plotfiddle{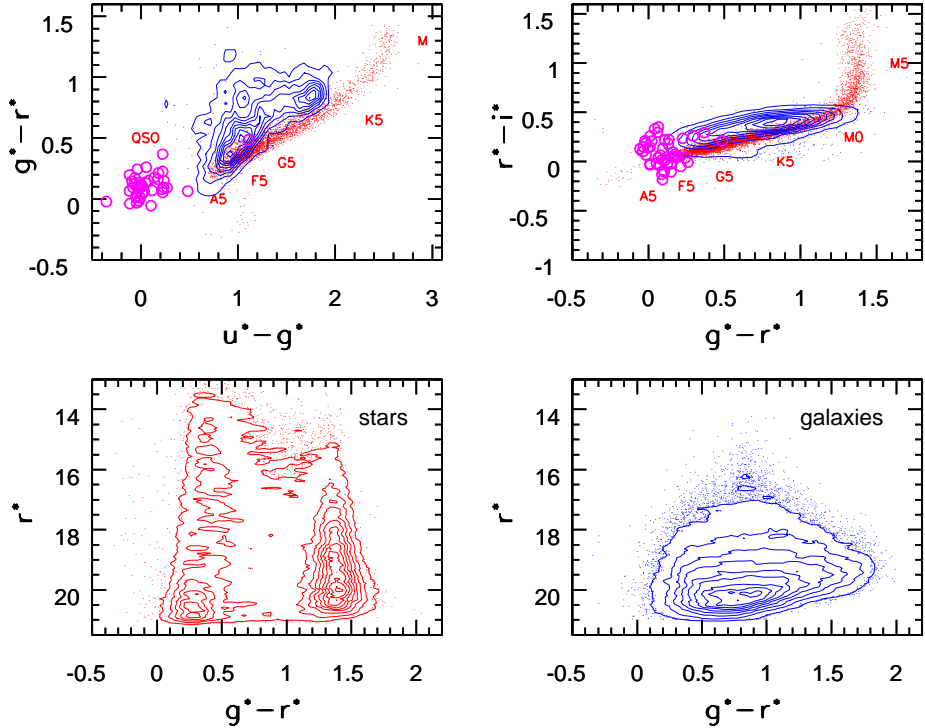}{9cm}{0}{65}{65}{-200}{-140}
\caption{The color-color and color-magnitude diagrams which summarize
photometric properties of SDSS sources.}
\end{figure}

\section{           SDSS Color-color and color-magnitude diagrams      }
\label{sdssccd}

Of the three surveys discussed here, SDSS provides the most detailed information 
about the detected objects due to its multi-color optical photometry, the highest
angular resolution, and the spectroscopic data. The position of an object in SDSS 
color-color and 
color-magnitude diagrams can be efficiently used to constrain its nature. The 
color-color and color-magnitude diagrams which summarize photometric properties 
of SDSS sources are shown in Figure 1 (magnitudes are marked as $m^*$ because
the calibration is still uncertain within \about5\%). We use the ``model" magnitudes, 
as computed by 
the photometric pipeline (``photo'', Lupton {\em et al.} 2002). The model magnitudes 
are measured by fitting an exponential and a de Vacouleurs profile, and using the 
formally better model in $r$ to evaluate the magnitude. Photometric errors are typically 
0.03$^m$ at the bright end ($r^* < 20^m$), and increase to about 0.1$^m$ at
$r^* \about 21^m$, the faint limit relevant in this work (for more details see 
Ivezi\'c {\em et al.} 2000 and Strateva {\em et al.} 2001, hereafter S01).

The top two panels in Figure 1 display the \gr\ vs. \ug\ and
\ri\ vs. \gr\ color-color diagrams for $\about$ 300,000 objects observed in 
20 deg$^2$ of sky. The unresolved sources are shown 
as dots, and the distribution of resolved sources is shown by linearly spaced density 
contours. The low-redshift quasars (z $\la$ 2.5), selected by their blue \ug\ colors 
indicating UV excess (0.6 $<$ \ug $<$ 0.6, -0.2 $<$ \gr $<$ 0.6), 
are shown as circles. Most of the unresolved sources marked as dots are stars.
For a more detailed discussion of the stellar properties in SDSS photometric system 
see Finlator {\em et al.} (2000, hereafter F00) and references therein. Here we only 
briefly mention  that the position of a star in color-color diagrams is mainly determined 
by its spectral type, as marked,  and that the modeling of the stellar 
populations observed by SDSS (F00) indicates that the majority of these stars (\about 
99\%) are on  the main sequence. The lower two panels in Figure 1 display the 
color-magnitude diagrams for unresolved (left) and resolved (right) sources, with 
the distributions shown by linearly spaced density contours. The distribution of 
galaxies in the SDSS color-color diagrams has been studied by Shimasaku {\em et al.} 
(2001) and S01. S01 found that galaxies show a strongly bimodal distribution of the 
\ur\ color, also visible in the upper left panel in Figure 1, and demonstrated that 
the two components can be associated with the spiral (blue component) and elliptical 
(red component) galaxies. 

The aim of this work is to find out where in the diagrams shown in Figure~1 2MASS
and FIRST sources are found, and how many 2MASS and FIRST sources are not detected by 
SDSS.

\begin{figure}[h]
\plotfiddle{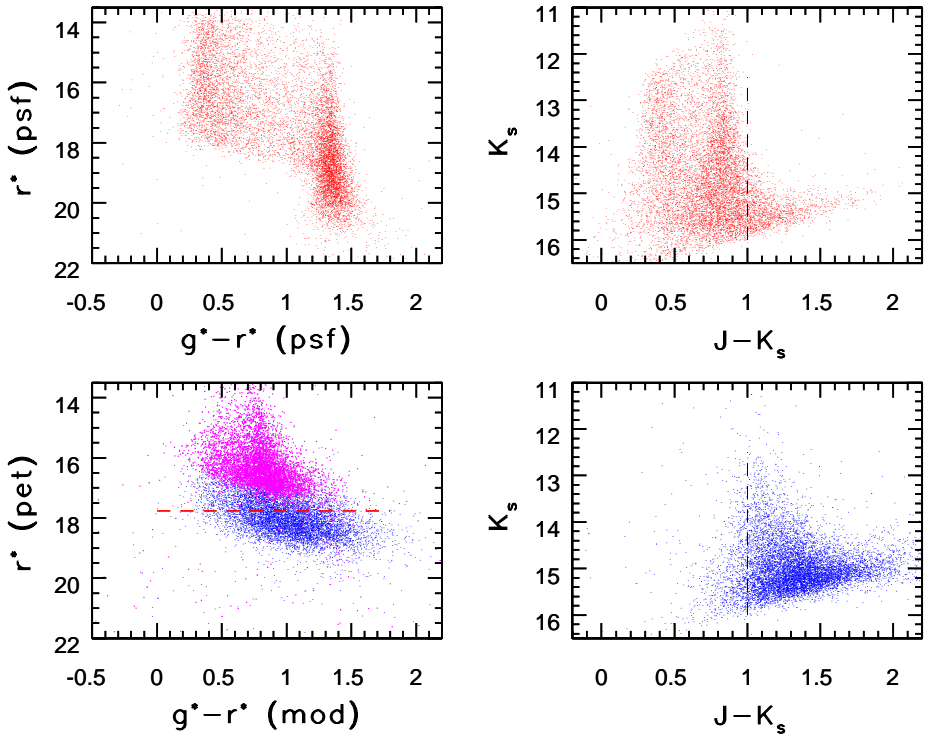}{9cm}{0}{65}{65}{-200}{-145} 
\caption{The color-magnitude diagrams for stars (top) and galaxies (bottom) observed 
by both SDSS and 2MASS.}
\end{figure}

\section{      The Positional Matching of SDSS and 2MASS Sources   }

\begin{figure}[t]
\plotfiddle{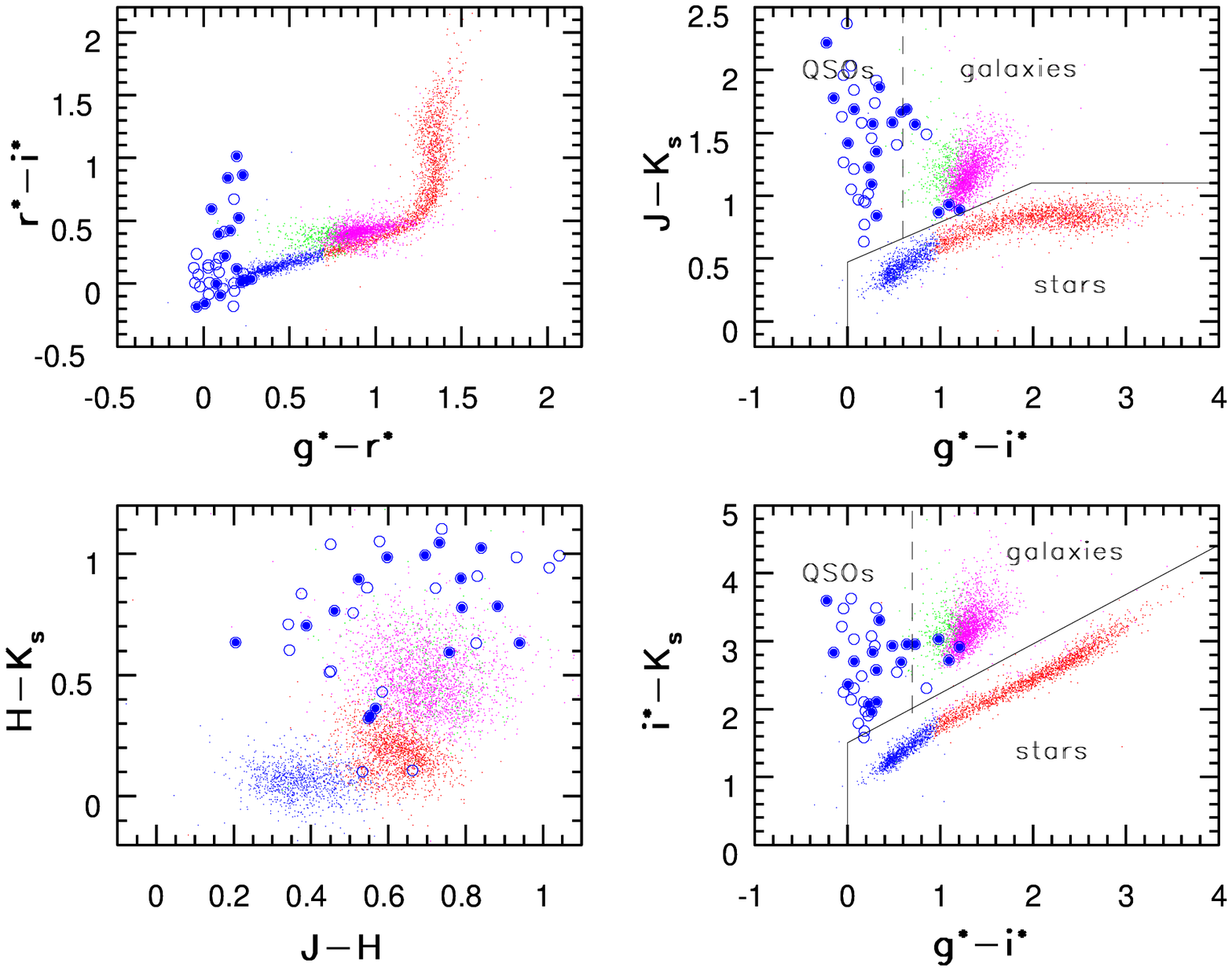}{9cm}{0}{65}{65}{-200}{-145} 
\caption{The color-color diagrams for sources from Figure 2 with high
signal-to-noise 2MASS detections.}
\end{figure}

\begin{table}[h]
\caption{The Galaxy Distribution$^a$ in the \r\ vs. \gr\ Diagram.}
\begin{tabular}{ccrrrrrrr}
& & & & & \\
\tableline
& Region &       Counts     &  Blue  &  Red  & PSC$^b$ & XSC$^b$ & FIRST \\
\tableline
&    Ia   &   24.6$\pm$0.5  &  95.0  &  5.0  &  38.6   &   17.9  &  3.9  \\ 
&    Ib   &   58.0$\pm$0.7  &  16.0  & 84.0  &  77.9   &   38.8  &  4.4  \\
&    Ic   &   13.2$\pm$0.3  &  2.2   & 97.8  &  79.1   &   20.9  &  8.8  \\
&   all I &   95.8$\pm$0.9  &  34.5  & 65.5  &  67.9   &   31.0  &  4.9  \\
&   IIa   &    317$\pm$1.8  &  97.4  &  2.6  &   1.9   &    0.0  &  0.1  \\
&   IIb   &    517$\pm$2.2  &  54.3  & 45.7  &   2.2   &    0.0  &  0.2  \\
&   IIc   &    304$\pm$1.8  &   7.4  & 92.6  &  14.4   &    0.0  &  0.8  \\
&   IId   &   84.0$\pm$0.9  &   2.4  & 97.6  &  21.0   &    0.1  &  3.3  \\
&  all II &   1222$\pm$3.5  &  50.2  & 49.8  &   6.5   &    0.0  &  0.5  \\
&  I + II &   1318$\pm$3.7  &  49.1  & 50.9  &  10.9   &    2.3  &  0.9  \\
\tableline
& & & & & \\
\end{tabular}
\leftline{a) All entries are percentages, except counts which are deg$^{-2}$.}
\leftline{b) Refers to 2MASS catalogs.}
\end{table}

\begin{figure}[t]
\plotfiddle{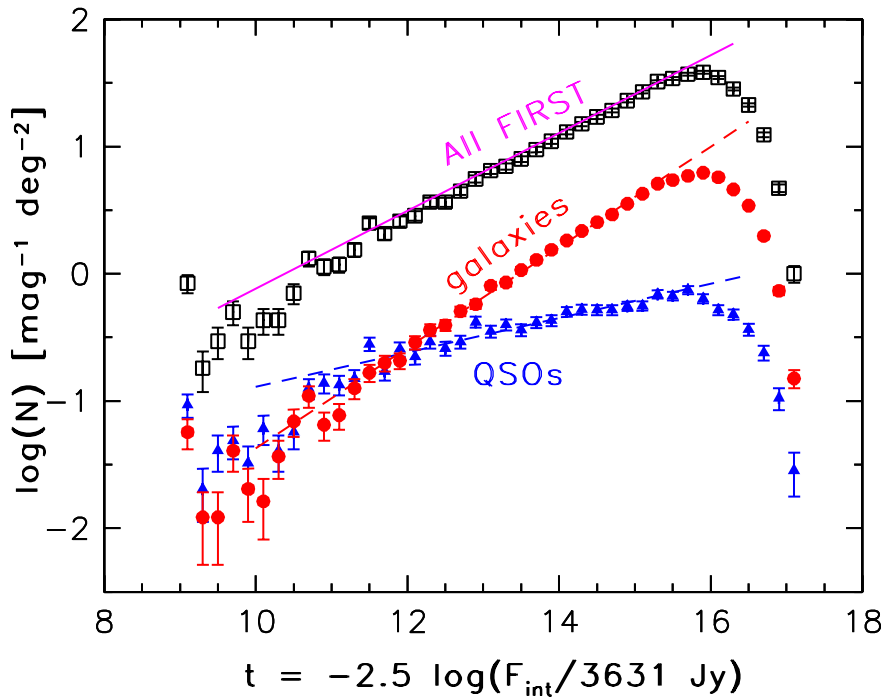}{8.5cm}{0}{75}{75}{-220}{-300} 
\caption{The comparison of the radio counts for SDSS-detected quasars 
and galaxies.}
\end{figure}

\begin{figure}[t]
\plotfiddle{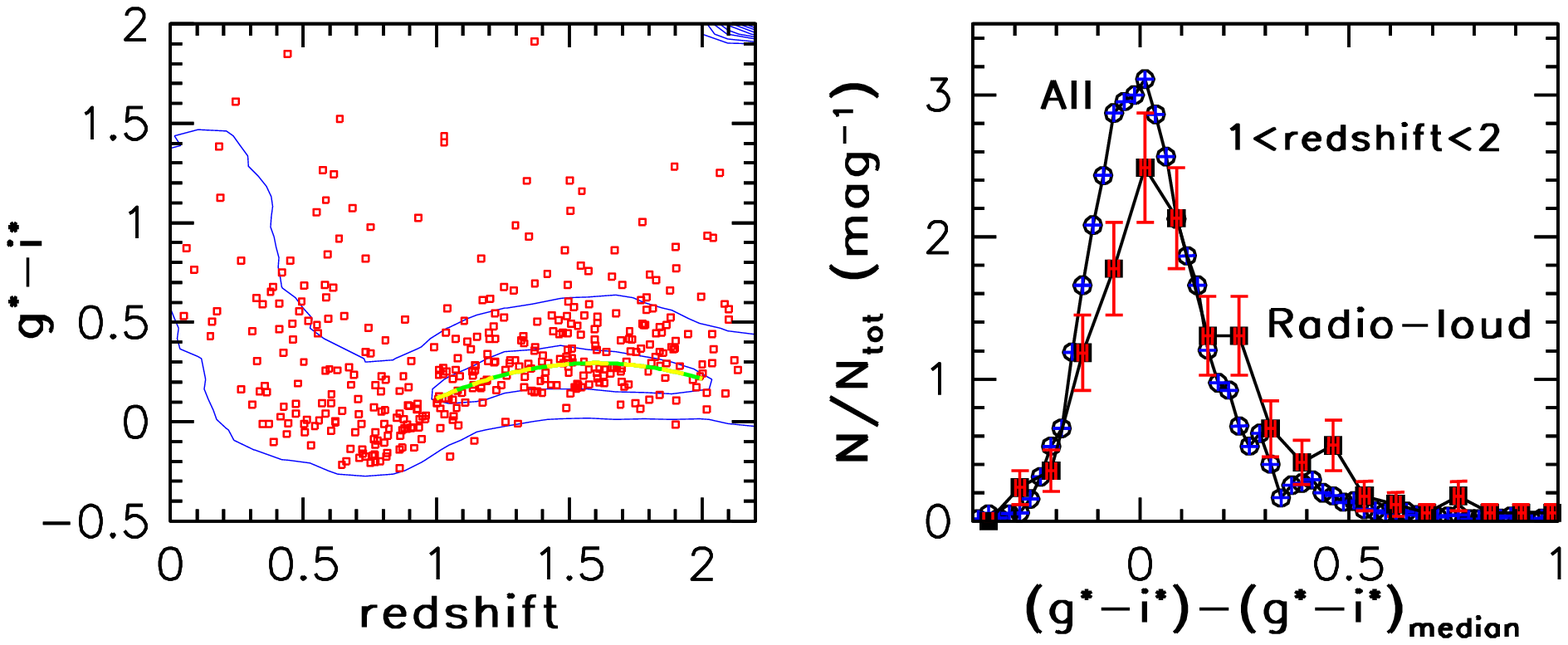}{5.0cm}{0}{70}{70}{-210}{-370}
\caption{The comparison of optical colors for radio-loud and optically
selected quasars (see Section 4.1).}
\end{figure}

The positional matching of SDSS and 2MASS sources is described 
by F00, who also discussed the optical and infrared properties of stars.
The analysis of extragalactic sources observed by both SDSS and 2MASS will
be described in detail by Ivezi\'c {\em et al.} (2002). Practically all 
2MASS sources ($\sim$98\% for point sources from the PSC and $\sim$97\%
for extended sources from the XSC) are matched to an SDSS source within 2 
arcsec. About $\sim$11\% of the 2MASS PSC sources are optically resolved 
galaxies and the rest are dominated by stars. Practically all ($\sim$98\%) 
sources from the 2MASS XSC are associated with an optically resolved source.
 
Figure 2 shows the color-magnitude diagrams for sources observed by both
SDSS and 2MASS. The top two panels show representative optical and near-IR diagrams
for \about 10,000 stars, and the bottom two panels show analogous diagrams for a 
similar number of galaxies. The vertical dashed lines at $J-K_s = 1$ in the two right 
panels roughly separate stars from galaxies in 2MASS data (only \about22\% of these 
galaxies are resolved in 2MASS images). The horizontal dashed line in the lower 
left panel shows the magnitude limit for SDSS spectroscopic galaxy survey. 
Practically all sources from the 2MASS XSC are brighter than that limit, and are 
shown as magenta dots (on top of blue dots that mark sources from the 2MASS PSC). 
The fraction of 2MASS galaxies in each of the regions in the \r\ vs. \gr\ 
color-magnitude diagram that track different morphological types (shown by dashed
lines in Figure 6, for details see Ivezi\'c {\em et al.} 2002) is listed in Table 1. 
Based on the matched source density, we project that by the completion of SDSS the 
matched SDSS-2MASS sample will include about 10$^7$ stars and 10$^6$ galaxies. 
About half of these galaxies will be part of the SDSS spectroscopic survey, and 
about about 10$^5$ will have photometry better than 0.1 mag. in all 8 bands.

Figure 3 shows the color-color diagrams for stars and galaxies from Figure 2, which
also have better than 10$\sigma$ detections in all three 2MASS bands. 
Stars are shown as blue and red dots, separated by \gr=0.7 which approximately 
corresponds to K0 spectral type (F00). Galaxies are shown as green and magenta dots, 
separated by $u^*-r^*=2.22$ which approximately separates spiral and elliptical 
galaxies (S01). The unresolved sources with optical colors indicative of 
low-redshift quasars are shown as blue circles (open for sources with better than 
5$\sigma$ detections in all three 2MASS bands and solid for better than 7$\sigma$).
Note that it is not possible to separate galaxies from stars using only optical 
colors (see upper left panel). However, the addition of IR data allows nearly perfect
separation, and the lines shown in the two right panels outline the regions populated 
by stars, galaxies and low-redshift quasars (the \i-$K_s$ color is based on the
``psf'' \i\ magnitude). This clean separation can be used to gauge
the success of the SDSS star-galaxy separation at the bright end.  We find that
99.3\% of the sources deemed as resolved by SDSS photometric pipeline are found
in the region marked as ``galaxies''. The bottom left panel shows an infrared
color-color diagram constructed with 2MASS data. Using SDSS data we find that 
sources with $H-K_s > 0.3$ are predominantly extragalactic, while the bluer
sources are stars.

\subsection{Search for Reddened Quasars Using SDSS and 2MASS}

Dust-reddened quasars are hard to distinguish in optical color-color diagrams 
because the plausible reddening vectors are roughly parallel to the stellar locus. 
However, such objects may be more easily distinguishable from stars in optical-infrared 
color-color diagrams because the reddening moves objects from the region marked 
``QSOs'' to the region marked ``galaxies'', thus missing the stellar locus 
(because quasars have redder $J-K_s$ colors than stars). Since SDSS imaging data 
can easily separate optically unresolved and resolved sources, selecting sources 
with e.g. $J-K_s>1$ and $\gi > 0.5$ can in principle reveal dust-obscured quasars. 
In a pilot study, we have obtained spectra for 30 optically unresolved sources 
from the region marked ``galaxies''. Most of them are M stars, and not a single 
one was confirmed to be a quasar. This null result places a strong upper limit on 
the fraction of reddened quasars.

Cutri {\em et al.} 2001 discuss the selection of red AGNs from the 2MASS PSC 
database using the condition that $J-K_s>2$. These sources may be dust-reddened
quasars with optical colors indistinguishable from those of stars and thus missed by 
SDSS spectroscopic targeting of quasars. We have analyzed a sample of  
241 such candidates selected from a 220 deg$^2$ large region (Ivezi\'{c},
Cutri, Nelson, {\em et al.} 2002). The majority of these sources are 
optically resolved (226), and are dominated by red (elliptical) galaxies (194) 
with redshifts up to \about0.4. Their very red $J-K_s$ colors seem to be 
a consequence of the K correction. The majority of optically unresolved matches 
(13 out of 15) show UV excess indicative of quasars with redshifts $\la$2.5 and
thus are easily distinguishable from stars using optical colors alone. This 
is confirmed for 9 sources for which SDSS spectra are available. One of the 
remaining two unresolved sources is a confirmed L dwarf, and the other one has 
optical-infared colors consistent with also being an L dwarf. Consequently, this 
preliminary analysis indicates that the fraction of moderately reddened quasars 
($A_V \la 5$) which are missed by SDSS quasar survey is small. While the precise 
value of this fraction is somewhat model-dependent, it seems to be less than 
\about10\%.

\section{      The Positional Matching of SDSS and FIRST Sources   }

The positional matching of SDSS and FIRST sources is described in detail
by Ivezi\'c {\em et al.} 2001 (see also Menou {\em et al.} 2001 and
Knapp {\em et al.} 2001). Optical identifications can be made for \about31\% of
FIRST sources. The majority of the FIRST sources identified with an SDSS source 
are optically resolved, and their fraction among the matched sources is a function 
of the radio flux, increasing from \about50\% at the bright end to \about90\% 
at the FIRST faint limit (1 mJy). The cumulative fraction of the optically
unresolved sources among the SDSS-FIRST matches is \about 16\%. The colors of 
optically unresolved sources, as well as  SDSS spectra for objects brighter 
than the flux cutoff for the spectroscopic targeting, indicate that they are 
dominated by quasars. Figure 4 compares the differential radio counts for 
SDSS-detected quasars, marked by dots, and galaxies, marked by triangles (we 
introduced an AB radio magnitude defined as  $t = -2.5 \log ({F_{int} /
3631 {\rm Jy}})$, where $F_{int}$ is the total radio flux). The dashed lines show 
the best fits in the 11.5 $< t < $ 15.5 range: for quasars
\eq{
               \log(N) = -2.24 + 0.14 \, t,
}
and for galaxies
\eq{
               \log(N) = -5.33 + 0.40 \, t.
}

We find no significant differences in the counts of FIRST sources with and without 
an optical identification: they both follow $\log(N) = C + 0.3 \, t$ relations. 
To illustrate this point, we compare the sum of counts for quasars and galaxies 
(renormalized to account for the matching fraction), shown as open squares, to the 
counts of all FIRST sources, shown by the solid line. As evident, the two  
distributions are similar.

Based on the matched source density, by the completion of SDSS and FIRST
the matched sample will include \about250,000 galaxies and \about50,000
quasars.  As discussed in detail by Ivezi\'c {\em et al.} 2001, the majority
of these quasars are radio-loud.

\subsection{  The Properties of Quasars Observed by SDSS and FIRST}

One of the most important advantages of a radio-selected sample of quasars is 
that it can be used to estimate the fraction of quasars with stellar colors that 
are missed by optical surveys, such as SDSS. The majority of optically unresolved 
SDSS-FIRST sources have non-stellar colors. We find that the fraction of sources 
with colors indistinguishable from stellar is \about10\%, assuming that the colors 
of radio-loud ($F_\nu^{radio} > 10 \, F_\nu^{optical}$, for more details see Ivezi\'c 
{\em et al.} 2001) quasars are similar to the colors of radio-quiet quasars. We do, 
however, detect a small but statistically significant difference between the colors 
of radio-loud quasars and radio-quiet quasars. The left panel in Figure 5 shows the 
dependence of QSO \g-\i\ color on redshift (for a similar dependence of other SDSS 
colors on redshift see Richards {\em et al.} 2001). The distribution of optically selected 
and unresolved quasars with \i $<$ 19 is shown by contours. The 464 radio-loud quasars 
from the matched sample are shown as squares. The dashed line shows the median \g-\i\ 
color of all optically selected quasars in the redshift range 1--2, which is subtracted 
from the \g-\i\ color to obtain a color excess. The right panel compares 
the distribution of this color excess for all optically selected quasars in that 
redshift range, shown by circles, to the distribution for 225 radio-loud quasars,
shown by squares. The \g-\i\ color-excess distribution for radio quasars appears to 
be different from the distribution for the whole sample. First, the median excess 
for the radio subsample is redder by \about0.05 mag, with about $4\sigma$ significance. 
Second, the fraction of objects with very large color excess is larger for the radio 
subsample. Adopting 0.5 mag for the minimum value of the color-excess, we find that 
2.4$\pm$0.2\% of quasars have such extreme \g-\i\ colors, while this fraction is 
7.1$\pm$1.8\% for the radio-loud quasars.

\subsection{  The Properties of Galaxies Observed by SDSS and FIRST}

\begin{figure}[t]
\plotfiddle{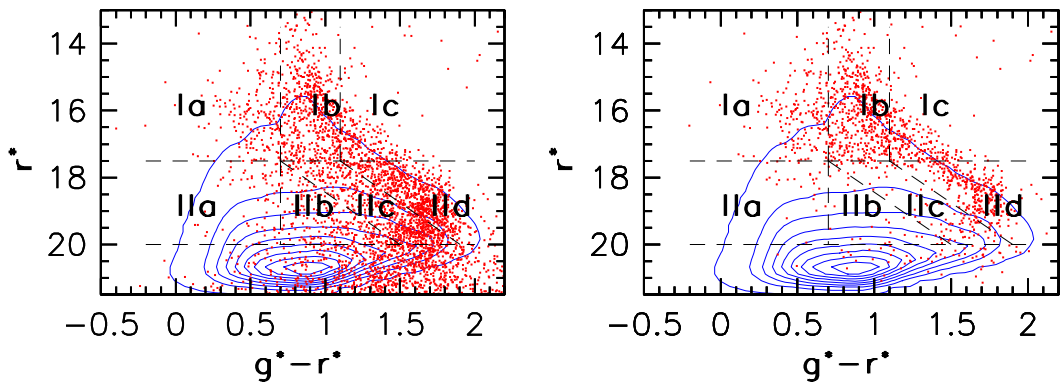}{5cm}{0}{75}{75}{-230}{-410}
\caption{The \r\ vs. \gr\ color-magnitude diagram for SDSS-FIRST galaxies. The left
panel shows all SDSS-FIRST galaxies, and the right panel those for which SDSS 
spectra are available.}
\end{figure}

Figure 6 displays the \r\ vs. \gr\ color-magnitude diagram for SDSS-FIRST galaxies,
shown by dots, compared to the distribution of all SDSS galaxies, shown by
linearly-spaced contours. The left panel shows the distribution of 
the SDSS-FIRST galaxies brighter than \r=21.5, and the right panel shows
the distribution of galaxies for which SDSS spectra are available. For 
details about the spectroscopic targeting of SDSS galaxies see 
Strauss {\em et al.} 2001 and Eisenstein {\em et al.} 2001. The dashed 
lines outline regions with different galaxy morphology and fraction of radio 
galaxies, as listed in Table 1. At the bright optical end (\r
$<$ 17.5) these radio-galaxies represent \about5\% of all SDSS galaxies, with the
radio fraction for red galaxies \about2 times higher than for blue galaxies.
In the magnitude range 17.5 $<$ \r $<$ 20, \about0.5\% of SDSS galaxies are detected 
by FIRST, and the radio fraction of the reddest galaxies, dominated by giant
ellipticals, is \about40 times larger than the radio fraction 
of the bluest galaxies. We find that radio galaxies in a redshift-limited sample 
have statistically indistinguishable colors and luminosity distribution from
other galaxies from the same volume. Nevertheless, the preliminary analysis of 
spectra indicates that the fraction of active galaxies is higher for the 
FIRST-detected galaxies than for all SDSS galaxies selected in the same
regions of the optical color-magnitude diagrams. A more quantitative analysis
of this effect will be presented in a future publication.

\section{ Discussion }

The preliminary analysis of sources in common to SDSS, 2MASS and FIRST
surveys indicates the enormous potential of combining the modern
digital sky surveys. The eight-color highly accurate photometry, the 
morphological information, radio properties, and redshift information
for several orders of magnitude larger samples than previously available
is bound to place the studies of extragalactic sources at an entirely 
new level. A good example of a result made possible by both a large sample
and accurate photometry is the small, yet statistically significant, 
difference between the optical colors of radio-loud and radio-quiet 
quasars. Another result with potentially large astrophysical significance
is the upper limit of 10\% on the fraction of quasars with stellar colors,
and finding that optical colors  of FIRST-detected galaxies are not 
significantly different from other galaxies in a volume-limited sample.   

\leftline{Acknowledgments}
{\small
The Sloan Digital Sky Survey (SDSS) is a joint project of The University of Chicago, Fermilab,
the Institute for Advanced Study, the Japan Participation Group, The Johns Hopkins University,
the Max-Planck-Institute for Astronomy (MPIA), the Max-Planck-Institute for Astrophysics
(MPA), New Mexico State University, Princeton University, the United States Naval
Observatory, and the University of Washington. Apache Point Observatory, site of the SDSS
telescopes, is operated by the Astrophysical Research Consortium (ARC). 
Funding for the project has been provided by the Alfred P. Sloan Foundation, the SDSS member
institutions, the National Aeronautics and Space Administration, the National Science
Foundation, the U.S. Department of Energy, the Japanese Monbukagakusho, and the Max Planck
Society. The SDSS Web site is http://www.sdss.org/.


}

\end{document}